\documentclass[a4paper]{article}
\usepackage{amsmath,graphicx,url,multirow}
\usepackage{bm,multirow,tabularx, threeparttable, amssymb}
\def\x{{\mathbf x}}

\def\x{{\bm x}}

\def\l{{\bm l}}
\def\bpi{{\bm \pi}}
\def\B{{\mathcal B}}
\def\bt{{\bm \theta}}

\usepackage{INTERSPEECH2020}

\title{CAT: A CTC-CRF based ASR Toolkit Bridging the Hybrid and the End-to-end Approaches towards Data Efficiency and Low Latency}
\name{Keyu An, Hongyu Xiang, Zhijian Ou$^{\dagger}$\thanks{$\dagger$ Corresponding author. This work is supported by NSFC 61976122. Code released at {https://github.com/thu-spmi/CAT}}}
\address{Speech Processing and Machine Intelligence (SPMI) Lab, Tsinghua University, China}
\email{aky19@mails.tsinghua.edu.cn, xianghy16@mails.tsinghua.edu.cn, ozj@tsinghua.edu.cn}

\begin{document}

\maketitle
\begin{abstract}
In this paper, we present a new open source toolkit for speech recognition, named CAT (\underline{C}TC-CRF based \underline{A}SR \underline{T}oolkit). CAT inherits the data-efficiency of the hybrid approach and the simplicity of the E2E approach, providing a full-fledged implementation of CTC-CRFs and complete training and testing scripts for a number of English and Chinese benchmarks. 
Experiments show CAT obtains state-of-the-art results, which are comparable to the fine-tuned hybrid models in Kaldi but with a much simpler training pipeline. Compared to existing non-modularized E2E models, CAT performs better on limited-scale datasets, demonstrating its data efficiency.
Furthermore, we propose a new method called contextualized soft forgetting, which enables CAT to do streaming ASR without accuracy degradation.
We hope CAT, especially the CTC-CRF based framework and software, will be of broad interest to the community, and can be further explored and improved.
\end{abstract}

\section{Introduction}

Deep neural networks (DNNs) of various architectures have become dominantly used in automatic speech recognition (ASR), which roughly can be classified into two approaches - the DNN-HMM hybrid and the end-to-end (E2E) approaches.
Initially, the DNN-HMM hybrid approach was adopted \cite{dahl2012context}, which is featured by using the frame-level loss (cross-entropy) to train the DNN to estimate the posterior probabilities of HMM states.
A GMM-HMM training is firstly needed to obtain frame-level alignments and then the DNN-HMM is trained.
The hybrid approach usually consists of an DNN-HMM based acoustic model (AM), a state-tying decision tree for context-dependent phone modeling, a pronunciation lexicon and a language model (LM), which can be compactly combined into a weighted finite-state transducer (WFST) \cite{mohri2008speech} for efficient decoding.

Recently, the E2E approach has emerged \cite{graves2006connectionist,Miao2015EESEN,graves2012sequence,chorowski2014end}, which is characterized by eliminating the construction of GMM-HMMs and phonetic decision-trees, training the DNN from scratch (in single-stage) and, even ambitiously, removing the need for a pronunciation lexicon and training the acoustic and language models jointly rather than separately.
The key to achieve this is to define a differentiable sequence-level loss of mapping the acoustic sequence to the label sequence.
Three widely-used E2E losses are based on Connectionist Temporal Classification (CTC) \cite{graves2006connectionist}, RNN-transducer (RNN-T)  \cite{graves2012sequence}, and attention based encoder-decoder \cite{chorowski2014end} respectively.

When comparing the hybrid and E2E approaches (modularity versus a single neural network, separate optimization versus joint optimization), it is worthwhile to note the pros and cons of each approach.
The E2E approach aims to subsume the acoustic, pronunciation, and language models into a single neural network and perform joint optimization.
This appealing feature comes at a cost, i.e. the E2E ASR systems are \emph{data hungry}, which require above thousands of hours of labeled speech to be competitive with the hybrid systems \cite{vs,chiu2018state,tuske2019advancing}.
In contrast, the modularity of the hybrid approach permits training the AM and LM independently and on different data sets. A decent acoustic model can be trained with around 100 hours of labeled speech whereas the LM can be trained on text-only data, which is available in vast amounts for many languages. In this sense, modularity promotes \emph{data efficiency}.
Due to the lack of modularity, it is difficult for an E2E model to exploit the text-only data, though there are recent efforts to alleviate this drawback \cite{toshniwal2018comparison,pham2019independent}.
In this paper, we are interested in bridging the hybrid and the E2E approaches, trying to inherit the data-efficiency of the hybrid approach and the simplicity of the E2E approach. 
A second motivation for such bridging is that low latency ASR has been addressed relatively easier and better in the hybrid approach than in the E2E approach, as will be discussed later in Section \ref{sec:related-work}.

Specifically, we base on the recently developed CTC-CRF approach \cite{Xiang2019CRF}. Basically, CTC-CRF is a CRF (Conditional Random Field) with CTC topology, which eliminates the conditional independence assumption in CTC and performs significantly better than CTC.
It has been shown \cite{Xiang2019CRF} that CTC-CRF has achieved state-of-the-art benchmarking performance with training data ranging from $\sim$100 to $\sim$1000 hours, while being end-to-end with a simplified pipeline (eliminating GMM-HMMs and phonetic decision-trees, training DNN-based AM in single-stage) and being data-efficient in the sense that cheaply available LMs can be leveraged effectively with or without a pronunciation lexicon.

In this paper we present CAT (\underline{C}TC-CRF based \underline{A}SR \underline{T}oolkit) towards data-efficient and low-latency E2E ASR, which trains CTC-CRF based AMs in single-stage and uses separate LMs, with or without a pronunciation lexicon.
On top of the previous work \cite{Xiang2019CRF}, the new contributions of this work are as follows.

\textbf{1.} 
CAT releases an full-fledged implementation of CTC-CRFs.
A non-trivial issue in training CTC-CRFs is that the gradient is the difference between empirical expectation and model expectation.
CAT contains efficient implementations of the forward-backward algorithm for calculating these expectations using CUDA C/C++ interface.
CAT adopts PyTorch \cite{paszke2017automatic} to build DNNs and do automatic gradient computation, and so inherits the power of PyTorch in handling DNNs.
In CAT, we can readily use the PyTorch DistributedDataParallel module to support training over multi-node and multi-GPU hardwares.

\textbf{2.} 
We add the support of streaming ASR in the toolkit.
To this end, we propose a new method called contextualized soft forgetting (CSF), which combines soft forgetting \cite{SForgetting} and  context-sensitive-chunk \cite{LC-BLSTM} in bidirectional LSTM (BLSTM).
Extensive experiments show that: (a) CTC-CRF with soft forgetting improves over CTC with soft forgetting significantly and consistently; (b) With contextualized soft forgetting, the chunk BLSTM based CTC-CRF with a latency\footnote{We define the latency as in \cite{SAN,CAS_transformer}, which is the time span corresponding to the right contextual frames. In our experiment, we use 10 right contextual frames by default, and the frames are computed with 10ms shift and 3-fold sampling.} of 300ms outperforms the whole-utterance BLSTM based CTC-CRF.

\textbf{3.} 
CAT provides reproducible, complete training and testing scripts for a number of English and Chinese benchmarks, including but not limited to WSJ, Switchboard, Fisher-Switchboard, and AISHELL datasets which are presented in this paper.
CAT achieves state-of-the-art ASR performance on these datasets, which are comparable to the LF-MMI \cite{povey2016purely} results in Kaldi (one of the strongest, fine-tuned hybrid ASR toolkit) but with a much simpler training pipeline. Remarkably, compared to existing non-modularized E2E models, CAT performs better on limited-scale datasets (with $\sim$100 to $\sim$2000 hours of training data), demonstrating its data efficiency.

\section{Related Work} \label{sec:related-work}

\textbf{~~~~ASR toolkits.} 
Roughly speaking, there are two approaches to using DNNs in ASR - the DNN-HMM hybrid and the E2E approaches. So does the classification of existing ASR toolkits.
For the hybrid approach, Kaldi \cite{Povey2012KALDI} may be the most widely-used hybrid DNN-HMM based ASR toolkit. In Kaldi, lattice-free maximum-mutual-information (LF-MMI) training needs a multi-stage pipeline consisting of GMM-HMM training and phonetic decision tree construction.
There have emerged some E2E ASR toolkits (e.g. ESPnet \cite{Watanabe2018ESPnet}/ESPRESSO \cite{wang2019espresso}, Wav2letter++ \cite{Pratap2018wav2letter}, and Lingvo \cite{shen2019lingvo}), mostly focusing on using attention-based encoder-decoder or hybrid CTC/attention.

EESEN \cite{Miao2015EESEN} and E2E-LF-MMI \cite{hadian2018end,SS-LF-MMI} seem to bridge the hybrid and the E2E approaches, by using the sequence-level loss (CTC and LF-MMI respectively) to train single-stage AMs and employing WFST based decoding.
EESEN is based on CTC, which, different from CTC-CRF, is limited by its conditional independence assumption and weak performance.
E2E-LF-MMI \cite{hadian2018end,SS-LF-MMI} was developed with two versions of using mono-phones or bi-phones, and bi-phone E2E-LF-MMI obtains comparable results to hybrid LF-MMI.
It is shown in our experiments that mono-phone CTC-CRF performs comparably to bi-phone E2E-LF-MMI but with a simpler pipeline. 
Bi-phone CTC-CRFs is found to slightly improve over mono-phone CTC-CRFs but will complicate the training pipeline.
%compared to bi-phone E2E-LF-MMI, mono-phone CTC-CRF performs equally over the (80-hour) WSJ dataset but better over the (260-hour) Switchboard dataset, with a simpler pipeline.
The differences between E2E-LF-MMI and CTC-CRF are detailed in \cite{Xiang2019CRF}.

\textbf{Low latency ASR.}
An important feature for a practical ASR toolkit is its ability to do streaming ASR with low latency.
In the hybrid approach, chunk-based schemes have been investigated in BLSTM \cite{LC-BLSTM,onlineNey}.
Time-delay neural networks (TDNNs) with interleaving LSTM layers (TDNN-LSTM) \cite{Low-latency} has been developed in Kaldi to successfully limit the latency while keeping the recognition accuracy.
In contrast, it is challenging and more complicated for attention-based encoder-decoders to do streaming ASR, which recently has received increasing studies, such as monotonic chunkwise attention (MoChA) \cite{MoChA}, triggered attention \cite{moritz2019triggered}, or using limited future context in the encoder \cite{CAS_transformer}. RNN-T has some advantage for streaming ASR but is data hungry, requiring large-scale training data to work. The RNN-T result over the Fisher-Swichboard data (2300 hours) \cite{transducer} is worse than CAT, as shown in Table \ref{Fisher-swbd}.

%A second issue sometimes overlooked for end-to-end ASR toolkits is the demand for low latency recognition which is crucial for streaming ASR applications.
%CTC-based Eesen uses bidirectional models by default.
%In attention based end-to-end systems (e.g. ESPnet), bidirectional encoder and global soft attention present inherent difficulty for low latency, and there are some works attempt to address it by providing limited future context to the encoder \cite{CAS_transformer,MERL_transformer}.
%There are recent efforts such as using monotonic chunkwise attention (MoChA) \cite{MoChA}, Latency-controlled BLSTM \cite{LC-BLSTM} and two-head cltLSTM \cite{ms_twohead}. 
%Online recognition with ESPnet has been recently studied \cite{is19_yan}.
%Wav2letter++ is based solely on convolutional neural networks, which use restricted future context and realize low latency. However, in order to model long-range dependencies, the neural network in \cite{Zeghidour2018Fully} is extremely deep and big (with 100 million parameters).

%Finally, for programming languages used in toolkits, Kaldi core primarily uses C++ which is efficient but not flexible in supporting various rapid developments in DNNs.
%PyTorch-Kaldi \cite{Ravanelli2018THE} builds the neural networks with PyTorch, and PyKaldi \cite{Pykaldi} allows users to interact with Kaldi and OpenFst via Python language.

\section{CTC-CRF based ASR}
\label{sec:Discriminative training with CRF}
CAT consists of separable AM and LM, which meets our rationale to be data efficient by keeping necessary modularity.
In the following we mainly describe our CTC-CRF based AMs.
CAT uses SRILM for LM training, and some code from Kaldi and EESEN for data preparation, decoding graph compiling and WFST based decoding. 
More details can be found in the toolkit.

Consider discriminative training of DNN-based AMs in single-stage based on the loss defined by conditional maximum likelihood \cite{Xiang2019CRF}:
\begin{equation} \label{eq:crf-obj1}
\mathcal{L}(\bt) = - \log p_{\bt}(\l|\x)
\end{equation}
where $\x \triangleq x_1,\cdots\, x_T$ is the speech feature sequence and $\l \triangleq l_1, \cdots\, l_L$ is the label (phone, character, word-piece and etc) sequence, and $\bt$ is the model parameter.
Note that $\x$ and $\l$ are in different lengths and usually not aligned.
To handle this, a hidden state sequence $\bpi \triangleq \pi_1,\cdots\,\pi_T$ is introduced; state topology refers to the state transition structure in $\bpi$, which basically 
defines a mapping $\mathcal{B}: S_\pi^{*} \to S_l^{*}$ that maps a state sequence $\bpi$ to a unique label sequence $\l$.
Here $S_l^{*}$ denote the set of all sequences over the alphabet $S_l$ of labels, and $S_\pi^{*}$ similarly for the alphabet $S_\pi$ of states.
It can be seen that HMM, CTC, and RNN-T implement different topologies.
CTC topology defines a mapping that removes consecutive repetitive labels and blanks, with $S_\pi$ defined by adding a special blank symbol $<$blk$>$ to $S_l$.
CTC topology is appealing, since it allows a minimum size of $S_\pi$ and avoids the inclusion of silence symbol, as discussed in \cite{Xiang2019CRF}.

Basically, CTC-CRF is a CRF with CTC topology. The posteriori of $\l$ is defined through the posteriori of $\bpi$ as follows:
\begin{equation} \label{eq:post-l} 
	p_{\bt}(\l | \x) = \sum_{\bpi \in \mathcal{B}^{-1}(\l)} p_{\bt}(\bpi | \x)
\end{equation}
And the posteriori of $\bpi$ is further defined by a CRF:
\begin{equation} \label{eq:post-pi}
p_{\bt}(\bpi|\x) = \frac{\exp(\phi_{\bt}(\bpi, \x))}{\sum_{\bpi'}{\exp(\phi_{\bt}({\bpi', \x}))}}
\end{equation}
Here $\phi_{\bt}(\bpi, \x)$ denotes the potential function of the CRF, defined as:
\begin{displaymath}
\phi_{\bt}(\bpi, \x) = \log p(\l)+ \sum_{t=1}^{T} \log p_{\bt}(\pi_t|\x)
\end{displaymath}
where $\l = \B(\bpi)$. 
$\sum_{t=1}^{T} \log p_{\bt}(\pi_t|\x)$ defines the node potential, calculated from the bottom DNN.
$\log p(\l)$ defines the edge potential, realized by an n-gram LM of labels and, for reasons to be clear in the following, referred to as the denominator n-gram LM.
Remarkably, regular CTC suffers from the conditional independence between the states in $\bpi$. In contrast, by incorporating $\log p(\l)$ into the potential function in CTC-CRF, this drawback is naturally avoided.
Combining Eq. (\ref{eq:crf-obj1})-(\ref{eq:post-pi}) yields the sequence-level loss used in CTC-CRF:
\begin{equation} \label{eq:crf-obj2}
\mathcal{L}(\bt) = - \log \frac{  \sum_{\bpi \in \mathcal{B}^{-1}(\l)} \exp(\phi_{\bt}(\bpi, \x))}{\sum_{\bpi'}{\exp(\phi_{\bt}({\bpi', \x}))}}
\end{equation}

The gradient of the above loss involves two gradients calculated from the numerator and denominator respectively, which essentially correspond to the two terms of empirical expectation and model expectation as commonly found in estimating CRFs.
Similarly to LF-MMI, both terms can be obtained via the forward-backward algorithm.
Specifically, the denominator calculation involves running the forward-backward algorithm over the denominator WFST $\bf{T}_{den}$.
$\bf{T}_{den}$ is an composition of the CTC topology WFST and the WFST representation of the n-gram LM of labels, which is called the denominator n-gram LM, to be differentiated from the word-level LM in decoding.

\section{Contextualized Soft Forgetting towards Streaming ASR}
\begin{figure}[t]
	\centering
	\centerline{\includegraphics[width=7.5cm]{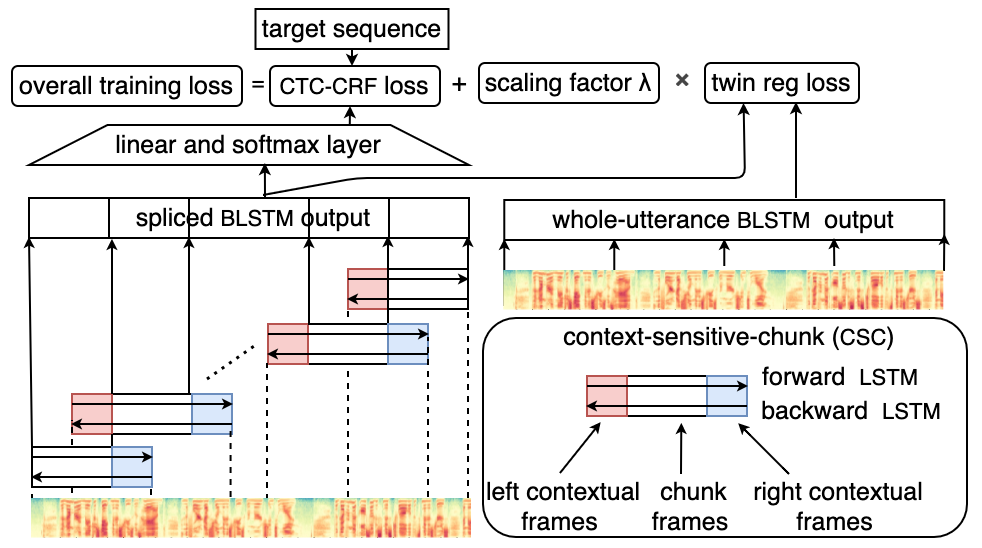}}
	%  \vspace{1.5cm}
	\caption{Contextualized Soft Forgetting for streaming ASR.}
	\label{CSF_fig}
		\vspace{-0.25cm}
\end{figure}
To enable streaming ASR in CAT, we draw inspirations from soft forgetting \cite{SForgetting} and context-sensitive-chunk \cite{LC-BLSTM} in using BLSTM.
With the hypothesis that whole-utterance unrolling of the BLSTM leads to overfitting, soft forgetting, which is developed for CTC-based ASR, consists of three elements. First, the BLSTM network is unrolled over non-overlapping chunks. The hidden and cell states are hence forgotten at chunk boundaries in training. Second, the chunk duration is perturbed across training minibatches, which is called chunk size jitter. Third, the CTC loss is added with a twin regularization term, which is the mean-squared error between the hidden states of a pre-trained fixed whole-utterance BLSTM and the chunk-based BLSTM being currently trained. Since twin regularization promotes some remembering across chunks, this method is called soft forgetting.
In streaming recognition, the hidden and cell states of the forward LSTM are copied over from one chunk to the next, and the backward LSTM hidden and cell states are reset to zero.

The idea of context-sensitive-chunk (CSC) is proposed in the BLSTM-HMM hybrid system to reduce the latency from a whole utterance to a chunk.
In CSC, a chunk is appended with a fixed number of left and right frames as left and right contexts.

In CAT, we propose to apply soft forgetting to context-sensitive-chunks, which is called contextualized soft forgetting (CSF) as illustrated in Figure \ref{CSF_fig}. 
First, we split an utterance into non-overlapping chunks.
For each chunk, a fixed number of frames to the left and right of the chunk are appended as contextual frames except for the first and last chunk, where we use zeros as the left and right contexts respectively. Thus we form context-sensitive-chunks and run BLSTM over each CSC.
The hidden and cell states of the forward and backward LSTM networks are reset to zeros at the left and right boundaries of each CSC in both training and inference.
When calculating the sequence-level loss in CTC-CRF, we splice the neural network output from chunks into a sequence again, but excluding the network outputs from contextual frames.
A pre-trained fixed whole-utterance BLSTM is used to regularize the hidden states of the CSC-based BLSTM, and the overall training loss is the sum of the CTC-CRF loss and the twin regularization loss with a scaling factor $\lambda$. 
Note that once the CSC-based BLSTM is trained, we can discard the whole-utterance BLSTM and perform inference over testing utterances without it.

\section{Experiment Settings}
The experiment consists of two parts. In the first part, we introduce the results on several representative benchmarks, including WSJ (80-h), AISHELL (170-h Chinese), Switchboard (260-h) and Fisher-Switchboard (2300-h) (the numbers in the parentheses are the size of training data in hours).
The performances over these limited-scale datasets reveal the data efficiency of different ASR models.
The second part presents the results for streaming ASR by the proposed contextualized soft forgetting method with ablation study.

It should be noted that the results shown in this paper should not be compared with results obtained with heavy data augmentation (e.g. specAugment \cite{specAugment}), much larger DNNs, and model combination.
When compared to results reported from other papers, unless otherwise stated, we cite those results under comparable conditions to the best of our knowledge.

\subsection{Setup for benchmarking experiment}
We compare CAT with state-of-the-art ASR systems on several benchmarks, as stated above. We apply speed perturbation for 3-fold training data augmentation, except on Fisher-Switchboard. Unless otherwise stated, 40 dimension filter bank with delta and delta-delta features are extracted. The features are normalized via mean subtraction and variance normalization per utterance, and sampled by a factor of 3.

The AM network, different from \cite{Xiang2019CRF}, is two blocks of VGG layers followed by a 6-layer BLSTM similar to \cite{Hori2017Advances}. We apply 1D max-pooling to the feature maps produced by VGG blocks on the frequency dimension only, since the input features have been sampled in time-domain and we find that max-pooling along the time dimension will deteriorate the performance. The first VGG block has 3 input channels corresponding to spectral features, delta, and delta delta features. The BLSTM has 320 hidden units per direction for WSJ and AISHELL, and 512 for Switchboard and Fisher-Switchboard. The total number of parameters is 16M and 37M respectively, much smaller than most E2E models. Denominator 4-gram LMs are used throughout the experiments. In training, a dropout \cite{Srivastava2014Dropout} probability of 50\% is applied to the LSTM to prevent overfitting. Following \cite{Xiang2019CRF}, a CTC loss with a weight $\alpha$ is combined with the CRF loss to help convergence. We set $\alpha$ = 0.01 by default and find in practice that the smaller $\alpha$ is, the better the final result will be.

\subsection{Setup for streaming ASR experiment}
To evaluate the effectiveness of contextualized soft forgetting, we first implement soft forgetting with the CTC-CRF loss. For a fair comparison, we adopt the same neural network architecture as in \cite{SForgetting}, which is a 6-layer BLSTM with 512 hidden units per direction. 40 dimension MFCC with delta and delta-delta are extracted, and the chunk size is set to 40. The whole-utterance BLSTM pre-trained on 260hr Switchboard obtain 14.3\% WER on eval2000. For twin regularization, the scaling factor $\lambda$ is set to 0.005. In contextualized soft forgetting, the chunk size is also 40, with 10 left and 10 right frames appended.

\section{Experimental results}
\subsection{Results for benchmarking experiment}

\begin{table}[th]
\vspace{-0.25cm}
	\centering
	\caption{Results over WSJ (80-h training data).}
		\vspace{-0.25cm}
	\scalebox{0.68}{
	\begin{tabular}{ccccc}
		\toprule
		\textbf{Model}    & \textbf{Unit}     & \textbf{LM}  & \textbf{dev93} & \textbf{eval92} \\
		\midrule
	    LF-MMI \cite{SS-LF-MMI} & mono-phone & 4-gram  & 6.0 & 3.0 \\
        LF-MMI \cite{SS-LF-MMI} & bi-phone & 4-gram  & 5.3 & 2.7 \\	
        \midrule
	    E2E-LF-MMI \cite{SS-LF-MMI} & mono-phone & 4-gram  & 6.3 & 3.1 \\
	    E2E-LF-MMI \cite{SS-LF-MMI}& bi-phone & 4-gram  & 6.0 & 3.0 \\
	
		EESEN \cite{Miao2015EESEN} & mono-phone & 3-gram  & 10.87& 7.28 \\

		ESPnet \cite{Watanabe2018ESPnet} & mono-char & RNN   & 12.4 & 8.9 \\
        Wav2letter++ \cite{Zeghidour2018Fully} & mono-char & 4-gram  & 9.5 & 5.6 \\
        Wav2letter++   \cite{Zeghidour2018Fully}& mono-char & Conv  & 7.5 & 4.1\\
        CTC/attention   \cite{Karita2019ACS}& mono-char & RNN  & 6.8 & 4.4\\
		\midrule
		CAT & mono-phone & 4-gram  & 5.7 & 3.2 \\
		CAT & mono-char & 4-gram  & 8.1 & 5.0 \\
		\bottomrule
	\end{tabular}}
\vspace{-0.25cm}
	\label{wsj}
\end{table}

\begin{table}[th]
%\vspace{-0.25cm}
	\centering
	\caption{Results over AISHELL (170-h Chinese training data).}
	\vspace{-0.25cm}
	  \scalebox{0.68}{
	  \begin{tabular}{cccc}
	  \toprule
	  \textbf{Model}    & \textbf{Unit}     & \textbf{LM}  & \textbf{Test}\\
	  \midrule
	  LF-MMI with i-vector \cite{Povey2012KALDI}  & tri-phone & 3-gram  & 7.43  \\
	  \midrule
      ESPnet \cite{Watanabe2018ESPnet}  & Chinese char & RNN  & 8.0  \\
      CTC/attention \cite{Karita2019ACS} & Chinese char & RNN & 6.7 \\
      Attention \cite{ACSAttention}  & Chinese char & RNN  & 18.7  \\
      Attention \cite{Component-Fusion} & Chinese char & RNN & 8.71 \\
	  \midrule
	  CAT & mono-phone & 3-gram & 6.34 \\
	  \bottomrule
	  \end{tabular}
	  }

\vspace{-0.35cm}
	\label{aishell}
\end{table}

\begin{table}[th]
\vspace{-0.4cm}
	\centering
	\caption{Results over Switchboard (260-h training data). The numbers in parentheses denote the results after rescoring with RNN-LMs. Results in square brackets denote the weighted average over SW and CH based on our calculation when not reported in the original paper. ``No LM'' denotes not using shallow fusion with external LMs.}
	\vspace{-0.25cm}
	\scalebox{0.68}{
			\begin{tabular}{cccccc}
				\toprule
				\textbf{Model}    & \textbf{Unit}     & \textbf{LM}  & \textbf{SW}  & \textbf{CH} & \textbf{Eval2000}\\
				\midrule
				LF-MMI \cite{SS-LF-MMI} & mono-phone & 4-gram  & 10.7 & 20.3 & [15.5]\\
				LF-MMI \cite{SS-LF-MMI} & bi-phone & 4-gram  & 9.5 (8.3) & 18.6 (17.1)&  [ 14.1 (12.7) ]\\
				\midrule
				E2E-LF-MMI \cite{SS-LF-MMI} & mono-phone & 4-gram  & 11.0 & 20.7 &[15.9]\\
				E2E-LF-MMI  \cite{SS-LF-MMI} & bi-phone & 4-gram  & 9.8 (8.5) & 19.3 (17.4) &[ 14.6 (13.0) ]\\
				EESEN \cite{Miao2015EESEN}  & mono-phone & 3-gram  & 14.8 & 26.0  &20.4\\
				Attention \cite{Zeyer2018RETURNN} & subword &No LM & 13.5 & 27.1  &20.3\\
				Attention \cite{zeyer2018improved} & subword & RNN  & 11.8 & 25.7 &18.1\\
				LAS \cite{specAugment} & subword &RNN  & 10.9  & 19.4 &[15.2]\\
				CTC/attention \cite{Karita2019ACS} & BPE & RNN & 9.0 & 18.1 & [13.6] \\
				\midrule
				
				CAT & mono-phone & 4-gram & 9.8 (8.8) & 18.8 (17.4) &14.3 (13.1)\\
				\bottomrule
			\end{tabular}
	}
\vspace{-0.25cm}
	\label{swbd}
\end{table}

\begin{table}[th]
\vspace{-0.1cm}
	\centering
	\caption{Results over Fisher-Switchboard (2300-h training data). Notations are the same as in Table \ref{swbd}.}
	  \vspace{-0.25cm}
	  \scalebox{0.68}{
	  \begin{tabular}{cccccc}
	  \toprule
	  \textbf{Model}    & \textbf{Unit}     & \textbf{LM}  & \textbf{SW} & \textbf{CH} & \textbf{Eval2000}\\
	  \midrule
	  LF-MMI \cite{SS-LF-MMI} & bi-phone & 4-gram  & 8.4 (7.5) & 15.1 (14.3) & [ 11.8 (10.9) ]\\
	  \midrule
E2E-LF-MMI \cite{SS-LF-MMI} & bi-phone & 4-gram  & 8.6 (7.6) & 15.4 (14.5) &[ 12.0 (11.1) ]\\
E2E-LF-MMI \cite{SS-LF-MMI} & mono-phone & 4-gram  & 8.9  & 16.8&[12.9]\\
RNN-T \cite{transducer} & char & 4-gram  & 8.1 & 17.5 &[12.8]\\
Attention \cite{improve-attention} & char & No LM & 8.3 & 15.5 &[11.9]\\
	  \midrule

	  CAT & mono-phone & 4-gram & 7.9 (7.3) & 16.0 (15.0) &12.0 (11.2)\\
	  \bottomrule
	  \end{tabular}}
	\label{Fisher-swbd}
\vspace{-0.25cm}
\end{table}
The WER results on WSJ are shown in Table \ref{wsj}, including two test sets - dev93 and eval92. 
It can be seen that CTC-CRF, hybrid LF-MMI and E2E-LF-MMI, which all keep modularity, perform comparable, and much better than other E2E models\footnote{Note that E2E models which use neural network based LMs via shallow fusion, are not directly compared to models using only n-gram LMs; they may be compared to models with RNN-LM rescoring.}. 

The CER (Character Error rate) results on AISHELL are shown in Table \ref{aishell}. 
%Note that pitch features are used in \cite{Povey2012KALDI,Watanabe2018ESPnet,ACSAttention} but not in CAT, because the pitch features are not suitable for composing a 3-channel feature map together with the fbank features. 
It can be seen that CTC-CRF obtains state-of-the-art performance on AISHELL dataset - the CER is much better than other E2E models and the hybrid LF-MMI in Kaldi.

The WER results on Switchboard are shown in Table \ref{swbd}. The Eval2000 test set consists of two subsets - Switchboard (SW) and Callhome (CH). 
It can be seen that compared to bi-phone hybrid LF-MMI and E2E-LF-MMI, mono-phone CTC-CRF performs comparably but with a simpler pipeline.
Remarkably, mono-phone CTC-CRF performs significantly better than other E2E models.

The WER results on Fisher-Switchboard are shown in Table \ref{Fisher-swbd}. The performance of CTC-CRF, with no data augmentation, is on par with state-of-the-art hybrid and E2E models.
Summing up the above results, we can see that on the limited-scale datasets (such as 80-h, 170-h, 260-h and 2300-h training data), the modularity of CTC-CRF clearly promotes data efficiency and achieve better results than other data hungry E2E models.

\subsection{Results for streaming ASR experiment}
\begin{table}[!th]
\vspace{-0.1cm}
	\centering
	\caption{Non-streaming recognition results for CTC and CTC-CRF, both trained with soft forgetting over (260-h) Switchboard. Non-streaming recognition means that the hidden and cell states of the forward and backward LSTMs are copied across chunk boundaries. Notations the same as in Table \ref{swbd}.}
	\vspace{-0.25cm}
	  \scalebox{0.8}{
	  \begin{tabular}{clccc}
	  \toprule
	  \textbf{Loss}  & \textbf{Model}    & \textbf{SW} & \textbf{CH} & \textbf{Eval2000}\\
	  \midrule
\multirow{3}{1cm}{\centering{CTC}}
	  &chunk-based \cite{SForgetting}&  12.7& 22.5 & [17.6]\\
	  & + chunk size jitter \cite{SForgetting}&  12.1& 21.5 & [16.8]\\
	  & + twin reg \cite{SForgetting}&  11.1& 19.7 & [15.4]\\
\midrule
\multirow{3}{1cm}{\centering{CTC-CRF}}
	   & chunk-based  &11.1  &19.6   & 15.4\\
	   &  + chunk size jitter & 10.5 & 18.8  & 14.7\\
	   & + twin reg & 10.0 & 18.8  & 14.4\\
	  \bottomrule
	  \end{tabular}
	  }

\vspace{-0.25cm}
	\label{SF}

\end{table}
First, we introduce different elements of soft forgetting \cite{SForgetting} in steps to show their impact on WERs and also compare CTC and CTC-CRF.
For this purpose, we follow \cite{SForgetting} to report the non-streaming recognition results, as shown in Table \ref{SF}.
We start from training the basic chunk-based BLSTM networks with a fixed chunk size.
It can be seen that CTC-CRF improves over CTC significantly under all experiment settings.

\begin{table}[!th]
\vspace{-0.1cm}
	\centering
	\caption{Streaming recognition results of CTC-CRFs, trained with Soft Forgetting (SF) and Contextualized Soft Forgetting (CSF) over (260-h) Switchboard.}
	\vspace{-0.25cm}
	\scalebox{0.8}{
			\begin{tabular}{clccc}
				\toprule
				\textbf{Method} & \textbf{Model}    & \textbf{SW} & \textbf{CH} & \textbf{Eval2000}\\
				\midrule
				\multirow{3}{1.6cm}{\centering{SF}}
				&chunk-based w/o future context &  11.0& 20.4 & 15.7\\
				&+ chunk size jitter &  11.0& 19.7 & 15.4\\
				&+ twin reg &  10.8& 19.7 & 15.3\\ 
				\midrule
				\multirow{3}{1.6cm}{\centering{CSF}}
				&chunk-based with context &  10.7& 20.0 & 15.4\\
				& + chunk size jitter &  10.4& 19.5 & 15.0\\
				& + twin reg &  9.7& 18.4 & 14.1\\
				\midrule
				\multirow{3}{1.6cm}{\centering{Results from literature}}
				& online-enabled BLSTM \cite{onlineNey} &  11.6& 23.0 & 17.3\\	
				& TDNN-D \cite{Low-latency} &  9.6& 19.9 & 14.8\\	
				& TDNN-LSTM-D \cite{Low-latency} &  9.0 & - & 13.9\\					
				\bottomrule
			\end{tabular}
	}

	\label{CSF}
\vspace{-0.35cm}
\end{table}

Then we examine the streaming recognition.
It can be seen from Table \ref{CSF} that CTC-CRFs trained with CSF improve significantly over CTC-CRFs with SF, and obtain comparable result with state-of-the-art TDNN-LSTM based hybrid model \cite{Low-latency}. 
Remarkably, the CSF based streaming CTC-CRF (14.1\%) even outperforms the whole-utterance CTC-CRF (14.3\%), presumably because CSF alleviates overfitting in addition to realizing streaming ASR.
This is in contrast to streaming ASR results by other E2E models, where streaming E2E models can hardly outperform their whole-utterance models \cite{SAN,CAS_transformer,MERL_transformer}.

\section{Conclusion}
This paper introduces an open source ASR toolkit - CAT, with the main features of data efficiency, simple pipeline, streaming ASR and superior results.
we propose a new method called contextualized soft forgetting, which enables CAT to do streaming ASR without accuracy degradation.
We hope CAT, especially the CTC-CRF based framework and software, will be of broad interest to the community, and can be further explored and improved, e.g. 
exploring different DNN architectures, different topologies of CRFs, and the application in more ASR tasks.
\bibliographystyle{IEEEtran}
\bibliography{mybib}

\end{document}